\documentclass[12pt]{article}

\usepackage[english]{babel}
\usepackage[utf8]{inputenc}
\usepackage[T1]{fontenc}

\usepackage{fullpage}
\usepackage{tabularx}
\usepackage{titlesec}
\usepackage{sectsty}
\usepackage{url}

\usepackage[dvipsnames]{xcolor}

\usepackage{booktabs}
\usepackage{breakcites}

\usepackage[colorlinks = true,urlcolor=BrickRed,linkbordercolor=red]{hyperref}
\usepackage{color}

\usepackage[labelfont=bf,font = small]{caption}
\usepackage{verbatim}
\usepackage{setspace}
\usepackage{caption}
\usepackage{abstract}
\usepackage{amsmath}
\usepackage{cleveref}
\usepackage{graphicx}

\title{Quantifying tissue growth, shape and collision via continuum models and Bayesian inference}

\author{Carles Falc\'o$^{1,}$\footnote{falcoigandia@maths.ox.ac.uk}\;, Daniel J. Cohen$^{2,3}$, Jos\'e A. Carrillo$^{1}$, Ruth E. Baker$^{1}$}
\date{}

\begin{document}

\maketitle
\vspace{-.75cm}
\begin{center}
{$^{1}$\emph{Mathematical Institute, University of Oxford, OX2 6GG Oxford, United Kingdom}}

{$^2$\emph{Department of Mechanical and Aerospace Engineering, Princeton University,\\Princeton, NJ, 08544, USA}}

{$^3$\emph{Department of Chemical and Biological Engineering, Princeton University,\\Princeton, NJ, 08544, USA}}
\end{center}

\begin{abstract}
   Although tissues are usually studied in isolation, this situation rarely occurs in biology, as cells, tissues, and organs, coexist and interact across scales to determine both shape and function. Here, we take a quantitative approach combining data from recent experiments, mathematical modelling, and Bayesian parameter inference, to describe the self-assembly of multiple epithelial sheets by growth and collision. We use two simple and well-studied continuum models, where cells move either randomly or following population pressure gradients. After suitable calibration, both models prove to be practically identifiable, and can reproduce the main features of single tissue expansions. However, our findings reveal that whenever tissue-tissue interactions become relevant, the random motion assumption can lead to unrealistic behaviour. Under this setting, a model accounting for population pressure from different cell populations is more appropriate and shows a better agreement with experimental measurements. Finally, we discuss how tissue shape and pressure affect multi-tissue collisions. Our work thus provides a systematic approach to quantify and predict complex tissue configurations with applications in the design of tissue composites and more generally in tissue engineering.
   \vspace{1cm}
\end{abstract}

\section{Introduction}

Cells do not live in isolation; instead they coexist and organize to form tissues and organs. In particular, during tissue growth, cells do not behave as isolated individuals, but sense their environment and direct their motion according to the information they receive. The sum of all individual cells, behaving in a coordinated manner and interacting with each other, can give rise to collective cell migration, which is essential for many different phenomena in biology, from wound healing and tumour invasion, to the formation of complex structures during development \cite{AlertTrepat,Schumacher2019}. Being such a fundamental process, much effort has been devoted to decipher the basic physical principles behind collective cell migration, both experimentally and from a modelling perspective \cite{chenpainter2020nonlocal, giniunaite2020modelling}. Being able to connect models and experimental data is thus essential in order to confirm the validity of mathematical models, as well as to gain further mechanistic insights.

At the tissue scale, mathematical models are usually based on a continuum description, where the cell density evolves according to a partial differential equation (PDE). Arguably the most famous continuum model of tissue spreading is the reaction-diffusion Fisher-KPP equation \cite{murray2001mathematical}, which is based on the assumption that cell movement is essentially random, and that cells proliferate according to a logistic growth law. This model and variants of it have been used to describe a variety of tissue formation experiments \cite{maini2004traveling,sengers2007experimental,sherratt1990models}.

From a biological perspective, however, the random motion assumption is not very realistic, as cells are able to sense the pressure exerted by neighbouring cells and direct their movement according to this information \cite{gurtin1977diffusion}. When population pressure is taken into account in continuum models one obtains the Porous-Fisher equation, which replaces the constant diffusion coefficient in the Fisher-KPP equation by a density-dependent function that increases as a power-law of the density. One of the most interesting features about this model is the appearance of compactly supported solutions, which give rise to the sharp invasion fronts observed in tissue formation experiments  \cite{browning2021model,CarrilloMurakawaCellAdhesion,el2021travelling,el2022continuum}. Of course, there are additional effects which can play an important role in collective cell motility and have been modelled using extensions of the mentioned equations, such as cell-cell adhesion \cite{ArmstrongPainterSherratt,domschke201441,falco2022local,MurakawaTogashi}, viscoelastic forces \cite{AlertTrepat,blanch2017effective,heinrich2020size},  interactions with the extracellular matrix \cite{browning2019bayesian,colson2021travelling,gerisch2008684}, heterogeneity in cell size \cite{khain2021dynamics,khain2018effective}, and cell-cycle dynamics \cite{simpson2020practical}.

Mathematical models can thus be more or less complex depending on the available data and the required level of biological detail, and they are a powerful tool to explore the impact of different biological mechanisms on collective cell movement. So-called \emph{identifiability analysis} methods \cite{chis2011structural,cobelli1980parameter} provide a systematic approach to relate model complexity to the type and amount of experimental data, and are a first step towards the estimation of model parameters. We say that a model is \emph{structurally identifiable} if different parameter values yield different model predictions. Hence, this is an intrinsic property of the model which depends on whether, given infinite ideal data, one can identify single values for the model parameters. Such formal structural identifiability analysis is possible for systems of ordinary differential equations \cite{janzen2016parameter,raue2014comparison}, and for certain families of PDEs (e.g. age-structured \cite{renardy2022structural}), but is more challenging for reaction-diffusion equations. Added to this, biological data is never infinite nor ideal which limits how much insight we can gain from structural identifiability.

As a result, here we explore the question of \emph{practical identifiability} \cite{chis2011structural} of two simple reaction-diffusion continuum models --- namely the Fisher-KPP and Porous-Fisher equations --- using data from recent tissue formation experiments
\cite{heinrich2020size,heinrich2021self}. Practical identifiability deals with finite and possibly noisy data, and depends on the inference method, but at its core is motivated by the same question: can we confidently identify estimates for the different model parameters? Here we follow the ideas in \cite{Hines2014DeterminationOP,simpson2020practical} and use a Bayesian approach in order to obtain posterior distributions for the different model parameters. Poor identifiability in a Bayesian context is thus associated with very broad posterior distributions indicating high uncertainty for the associated parameters \cite{siekmann2012mcmc}.

Our work reveals that both models can be suitably calibrated to reproduce the dynamics of freely expanding epithelia, with the different model parameters being practically identifiable in all considered settings. However, when tissues are not isolated from each other and are allowed to collide as a result of motility and proliferation during tissue growth, only the Porous-Fisher model, which considers interactions between cells, is able to describe the experimental data. This model, while being relatively simple and having only three parameters, also proves to be useful for understanding the dynamics of multi-tissue collisions and, hence, for predicting steady state tissue configurations with applications in tissue engineering.

We structure the paper as follows; first, we describe the two continuum models and the inference approach taken. Then, we estimate the different model parameters using comprehensive experimental data of the growth of large, circular epithelia (Heinrich et al. \cite{heinrich2020size}). After confirming that the two employed models are practically identifiable and that they can reproduce data collected from these experiments, we validate our models on more complex experimental datasets detailing how multiple epithelia interact with each other during collision and healing experiments (Heinrich et al. \cite{heinrich2021self}). Using the obtained parameter estimates we explore whether the two models can reproduce several tissue collisions experiments with very different initial tissue geometries. Finally, we use the Porous-Fisher model to quantify and characterize the dynamics of multi-tissue collisions.

\enlargethispage{0.15cm}

\section{Simple models of tissue growth}

We start by looking at simple models describing the growth of a single epithelial monolayer tissue. We denote cell density in the tissue by a continuous variable $\rho(\mathbf{x},t)$ which depends on space $\mathbf{x}$ and time $t$. Cell density is assume to change due to cell movement and local proliferation. Mass conservation implies then that the density $\rho$ satisfies the continuity equation
\begin{equation}
\label{eq:continuity_eq}
    \partial_t\rho +\nabla\cdot\mathbf{j}  = r \rho f(\rho),
\end{equation}
where the flux $\mathbf{j}$ determines how cells move, $r$ is the proliferation rate, and $f(\rho)$ is a crowding function which regulates how density-dependent effects reduce net growth. For simplicity we consider logistic growth given by $f(\rho) = 1 - \rho/K$, with $K$ a saturation density or carrying capacity. Note that epithelial tissues are well characterized to undergo contact inhibition of proliferation, where cell cycling decreases as cell density increases \cite{heinrich2020size,puliafito2012collective} and hence the logistic growth assumption is reasonable --- see also \cite{SIMPSON2022110998} for other possibilities.

A very simple model can be motivated by assuming that cells move randomly following Brownian motion, which corresponds to the well-known Fick's law of diffusion, $\mathbf{j} = -D\nabla\rho$. In this case, we obtain the Fisher-KPP equation
\begin{equation}
    \partial_t\rho = D\Delta\rho+r\rho\left(1 - \frac{\rho}{K}\right).
    \label{eq:linear_diffusion_1species}
\end{equation}
This model and related ones are particularly relevant to describe tissue growth due to the presence of travelling wave solutions, which are characterized by an invasion front of fixed shape that propagates at a constant speed \cite{murray2001mathematical}.

However, a more realistic model should account for the fact that cell movement is not completely random and can be influenced by the local cell density. A standard approach in order to incorporate crowding effects into Eq. \eqref{eq:continuity_eq} results from the assumption that the velocity is proportional to the gradient of the  density, so that cells move down population density gradients. In other words, we write the flux as $\mathbf{j} = \rho\mathbf{v}$, where $\mathbf{v}$ represents the cell velocity and now assume that $\mathbf{v} = -D\nabla \rho$. This gives the following Porous-Fisher equation
\begin{equation}
    \partial_t\rho = D\,\nabla\cdot\left(\rho\nabla\rho\right)+r\rho\left(1 - \frac{\rho}{K}\right).
    \label{eq:nonlinear_diffusion_1species}
\end{equation}

When there is no proliferation ($r=0$), Eq.~\eqref{eq:nonlinear_diffusion_1species} corresponds to a specific case of the well-known porous-medium equation \cite{PME}. This equation is also related to Darcy's law which links the velocity with the population pressure: $\mathbf{v} = -\nabla P(\rho)$. For the general porous-medium equation, pressure and density are related via the power-law function $P(\rho)\sim \rho^{m-1}$, depending on the exponent $m$. During this work and unless stated otherwise, we will assume $m = 2$. Note that in the limit $m\rightarrow 1$, one obtains the linear diffusion case with $P(\rho)\sim\log\rho$.

From a microscopic point of view, where one focuses on individual cell trajectories, Eq.~\eqref{eq:linear_diffusion_1species} corresponds to the continuum limit of a system of non-interacting agents which move randomly and can proliferate with a density-dependent probability. The porous-medium equation with $m = 2$ can also be derived from microscopic movement rules when one takes into account volume exclusion \cite{calvez2006volume,gamba2003,painter2002volume}, starting from on-lattice \cite{bakerAspectRation,falco2022random} and also from off-lattice agent-based models \cite{CarrilloMurakawaCellAdhesion,DysonVolumeExclusion,DysonMacroscopicCrowding,Oelschlger1990LargeSO}. Further, the case with $m=3$ can be identified as the mean-field limit of a system of interacting agents with a particular diffusive scaling \cite{worsfold2022density} and has also been suggested as \emph{the simplest model} to relate the dispersal velocity to both the density and its gradient \cite{TopazBertozziLewis}.

In the following, we connect Eqs.~\eqref{eq:linear_diffusion_1species} and~\eqref{eq:nonlinear_diffusion_1species} with data from recent experiments studying the dynamics of expanding and colliding epithelial monolayer tissues. The two suggested models are solved numerically in two spatial dimensions with the finite-volume numerical scheme described in \cite{nonlocalNumericalSchemeRafaMarkus,nonlocalSchemceCarrilloYanghong}.

\section{Single tissue expansions and parameter estimation}

\begin{figure}
    \centering
    \includegraphics[width = \textwidth]{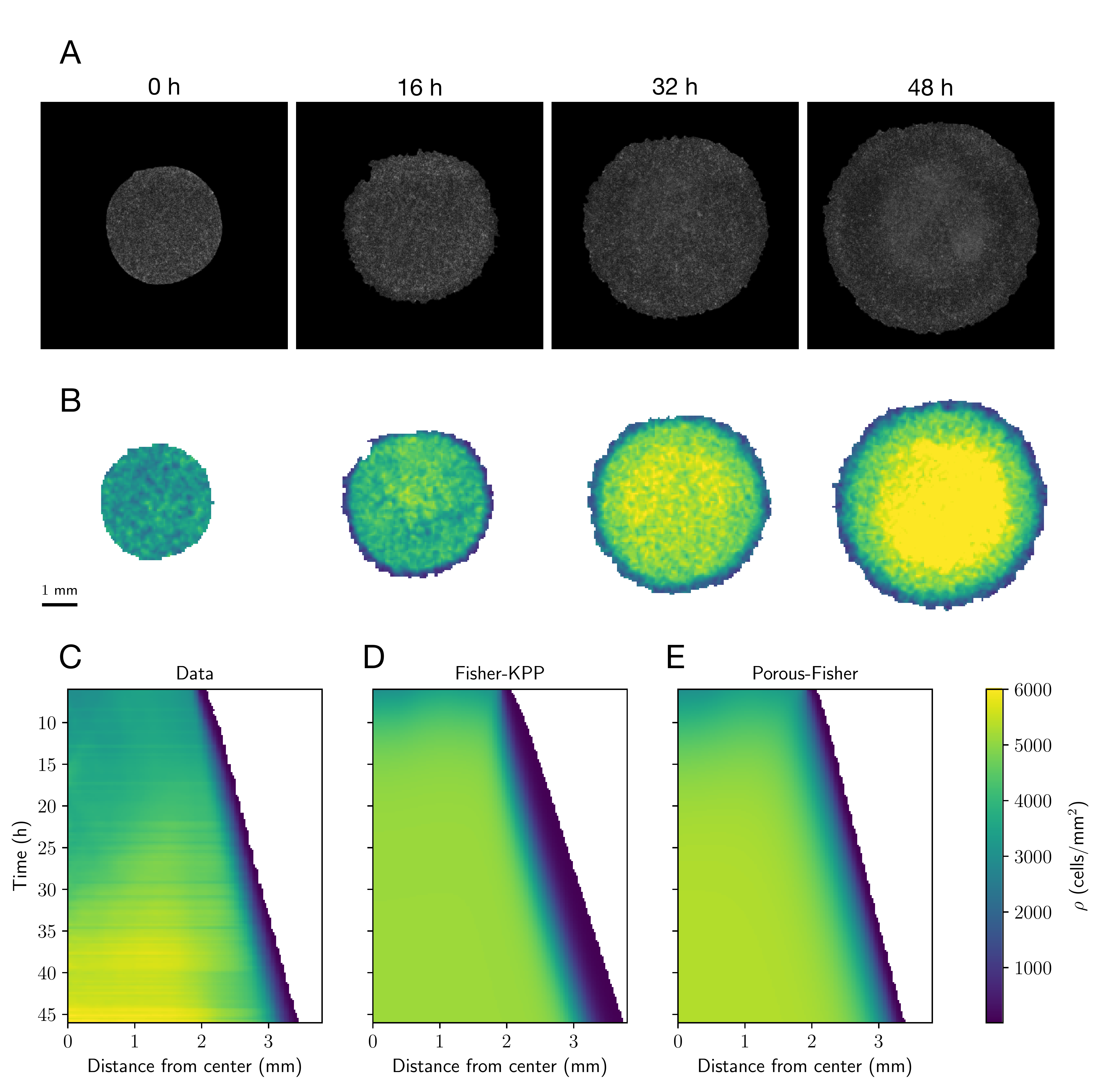}
    \caption{Expansions of single tissues and model predictions. (A) Microscopy images in phase-contrast at different times for the expansion of a circular tissue with initial diameter of 3.4 mm --- taken from \cite{matthew_a_heinrich_2020_3858845}. (B) Quantified experimental cell densities for the same expansion --- data from \cite{matthew_a_heinrich_2020_3858845}. (C) Experimental radial density profile obtained after averaging the expansions of 11 tissues with the same initial condition --- from \cite{heinrich2020size}. (D) Radial density profile from the Fisher-KPP model given by Eq.~\eqref{eq:linear_diffusion_1species}. (E) Radial density profile from the Porous-Fisher model given by Eq.~\eqref{eq:nonlinear_diffusion_1species}. Model parameters correspond to the maximum posterior estimates. All densities thresholded at 10 cells/mm$^2$. See Figure S3 for individual density profiles at specific time points}
    \label{fig:expansions}
\end{figure}

In order to calibrate the two suggested models, we focus on the experiments by \cite{heinrich2020size}. In these, Heinrich et al. characterized the expansion dynamics and growth of single circular epithelial tissues using an MDCK cell line. Initially, cells are cultured in a silicone stencil for 18 hours and, after the stencil removal, tissues are allowed to freely expand for 46 hours, which enables for each cell to undergo 2-3 cell divisions given that the cell cycle duration is around 16 hours. Local densities are then quantified by counting the number of nucleus centroids --- for more details we refer to \cite{heinrich2020size}. For our analysis, we only consider the measured cell densities after the first six hours of the experiment so that effects caused by the stencil removal are negligible. In Figure \ref{fig:expansions}A we show snapshots from one such experiment using a circular tissue with initial diamater of 3.4 mm --- see Figure \ref{fig:expansions}B for the quantified densities. The radial density profile resulting from averaging 11 experimental replicates is shown in Figure \ref{fig:expansions}C. Datasets used to reproduce these figures were taken from \cite{matthew_a_heinrich_2020_3858845}. See Figure S3 for individual density profiles at specific time points.

Our experimental data then consists of many individual measurements of the cell densities $\rho(\mathbf{x},t)$, giving rise to the dataset $\mathcal{D}=\{\rho^\mathcal{D}(\mathbf{x}_i,t_j)\}_{i,j}$. Here, the different measurements are recorded every 20 minutes, while the positions $\{\mathbf{x}_i\}_i$ correspond to the centers of small voxels of $115\times 115$ $\mu$m$^2$. In practice, and in order to keep the dimensionality of the data sufficiently low, we will only use the densities corresponding to the time points $t_j = 16,26,36,46$ h. In order to connect experimental data and models, we assume that the observations $\rho^\mathcal{D}$ are noisy versions of the model predicted density $\rho$. A common approach in mathematical biology \cite{Hines2014DeterminationOP,simpson2020practical} is to impose that the observation errors are additive, independent and normally distributed with variance $\sigma^2$. In other words, we assume the following error model
\begin{equation}
    \rho^\mathcal{D}(\mathbf{x}_i,t_j) = \rho(\mathbf{x}_i,t_j) + \varepsilon,\quad \varepsilon\sim \mathcal{N}(0,\sigma^2).
    \label{eq:error_model}
\end{equation}

\subsection{Parameter estimation via maximum likelihood}

Both models --- Eqs.~\eqref{eq:linear_diffusion_1species} and~\eqref{eq:nonlinear_diffusion_1species} --- have three parameters $D,r,K$ to be estimated. Considering the variance of the observation error as an extra parameter, we can write them as a vector $\theta = \left(D,r,K,\sigma\right)$. With the error model given by Eq.~\eqref{eq:error_model} we can explicitly write the log-likelihood of observing the measured data
\begin{equation}
    \ell(\theta) = -\frac{1}{2}\sum_{i,j}\left(\log\left({2\pi\sigma}\right)-\left(\frac{\rho(\mathbf{x_i},t_j) - \rho^\mathcal{D}(\mathbf{x_i},t_j)}{\sigma}\right)^2\right).
    \label{eq:log_likelihood}
\end{equation}
A direct approach to estimating the parameters in the two models consists of  maximixing this log-likelihood as a function of the parameter vector $\theta$, which gives a maximum likelihood estimator of the model parameters: $\theta_{\text{ML}} = {\mathrm{argmax}}_{\theta}\,\ell(\theta)$. In the case of a fixed noise parameter $\sigma$, this is equivalent to minimizing the 2-norm of the difference between model and data. Note, however, that whenever the models are non-identifiable maximizing the likelihood might lead to misleading results \cite{siekmann2012mcmc}. This is thus only a first step in our parameter inference analysis.

We perform the likelihood optimization using the parameter inference toolbox pyPESTO  \cite{pypesto}. This toolbox allows for local optimization of the likelihood starting from an initial guess of $\theta_0$. By randomly sampling a large number of initial vectors $\theta_0$ we find the same local maximum in most of the optimization runs. Additionally, this local maximum also maximizes the likelihood among all the found local maxima. In order to generate initial guesses of $\theta_0$ we sampled uniformly on log-scale using the parameter bounds $10^{-2.5}<r<10^1$ h$^{-1}$, $10^3<K<10^{3.5}$ cells/mm$^2$, $10^1<\sigma<10^{3.5}$ cells/mm$^2$, for both models; and $10^{2.5}<D<10^{4.5}$ $\mu$m$^2$/h for the Fisher-KPP model, and $10^{-1.5}<D<10^{-2.5}$ $\mu$m$^2$/h, for the Porous-Fisher model. The maximum likelihood estimators are indicated using dashed lines in Figures \ref{fig:posteriors_linear} and \ref{fig:posteriors_nonlinear} for the Fisher-KPP and Porous-Fisher models respectively.

As stated earlier, only the experimental cell densities corresponding to the time points $t_j = 16,26,36,46$ h were used for the likelihood calculation in Eq.~\eqref{eq:log_likelihood}. This was done to keep the computational costs of computing the maximum likelihood estimate at reasonable levels. Different choices of these time points yielded similar results for the maximum likelihood estimate $\theta_{\text{ML}}$.

\subsection{Bayesian inference}

 Here we explore the question of \emph{practical identifiability} of the Fisher-KPP and Porous-Fisher models using a Bayesian approach. In order to capture uncertainty in the model parameters, we are interested in estimating the posterior distribution $P(\theta\,| \,\rho^\mathcal{D})$, which can be calculated from Bayes' theorem
\begin{equation*}
    P(\theta\,| \,\rho^\mathcal{D})\propto P(\rho^\mathcal{D}\,| \,\theta)\pi(\theta),
\end{equation*}
where $P(\rho^\mathcal{D}\,| \,\theta)$ is the likelihood of observing the measured data, and $\pi(\theta)$ is the prior distribution of the parameter vector $\theta$. We assume the error model given by Eq.~\eqref{eq:error_model}, and hence the log-likelihood is given by Eq.~\eqref{eq:log_likelihood}. The priors for the two considered models are assumed to be uniform on log-scale using the bounds given in the previous section.

\begin{figure}[t]
    \centering
    \includegraphics[width = \textwidth]{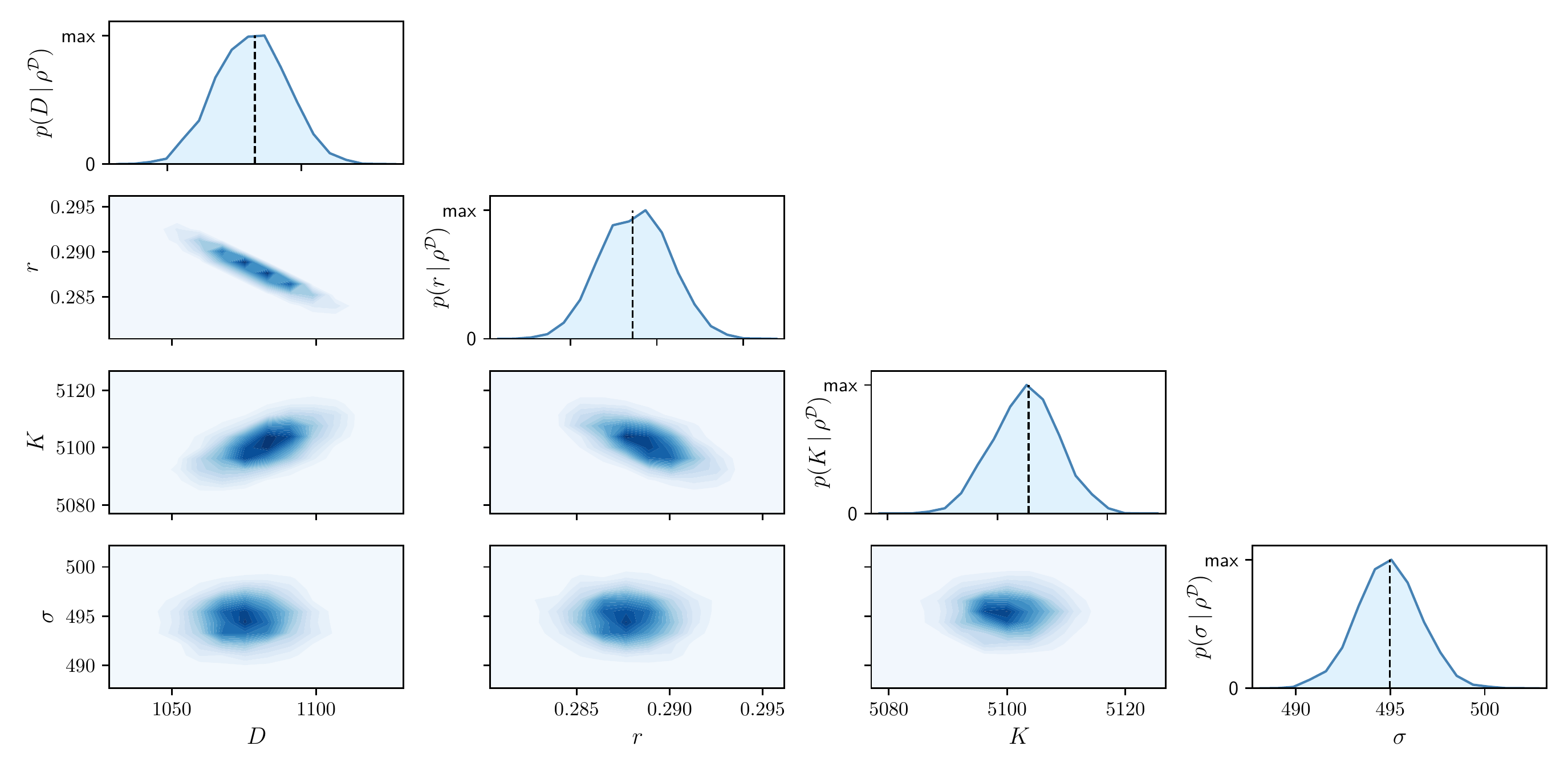}
    \caption{Results of the MCMC algorithm for the Fisher-KPP model given by Eq.~\eqref{eq:linear_diffusion_1species}. The diagonal plots represent the univariate marginal posterior distributions for each parameter. Below the diagonal we show the bivariate densities for every combination of parameters. Univariate posterior modes correspond to $(D,r,K,\sigma) = (1073\pm 13$ $\mu$m$^2$/h, $0.289\pm0.002$ h$^{-1}$, $5113\pm6$ cells/mm$^2$, $492\pm2$ cells/mm$^2$), where the errors are given by one standard deviation, calculated from the posterior distributions. Black dashed lines indicate the maximum likelihood estimates for each parameter.}
    \label{fig:posteriors_linear}
\end{figure}

\begin{figure}[t]
    \centering
    \includegraphics[width = \textwidth]{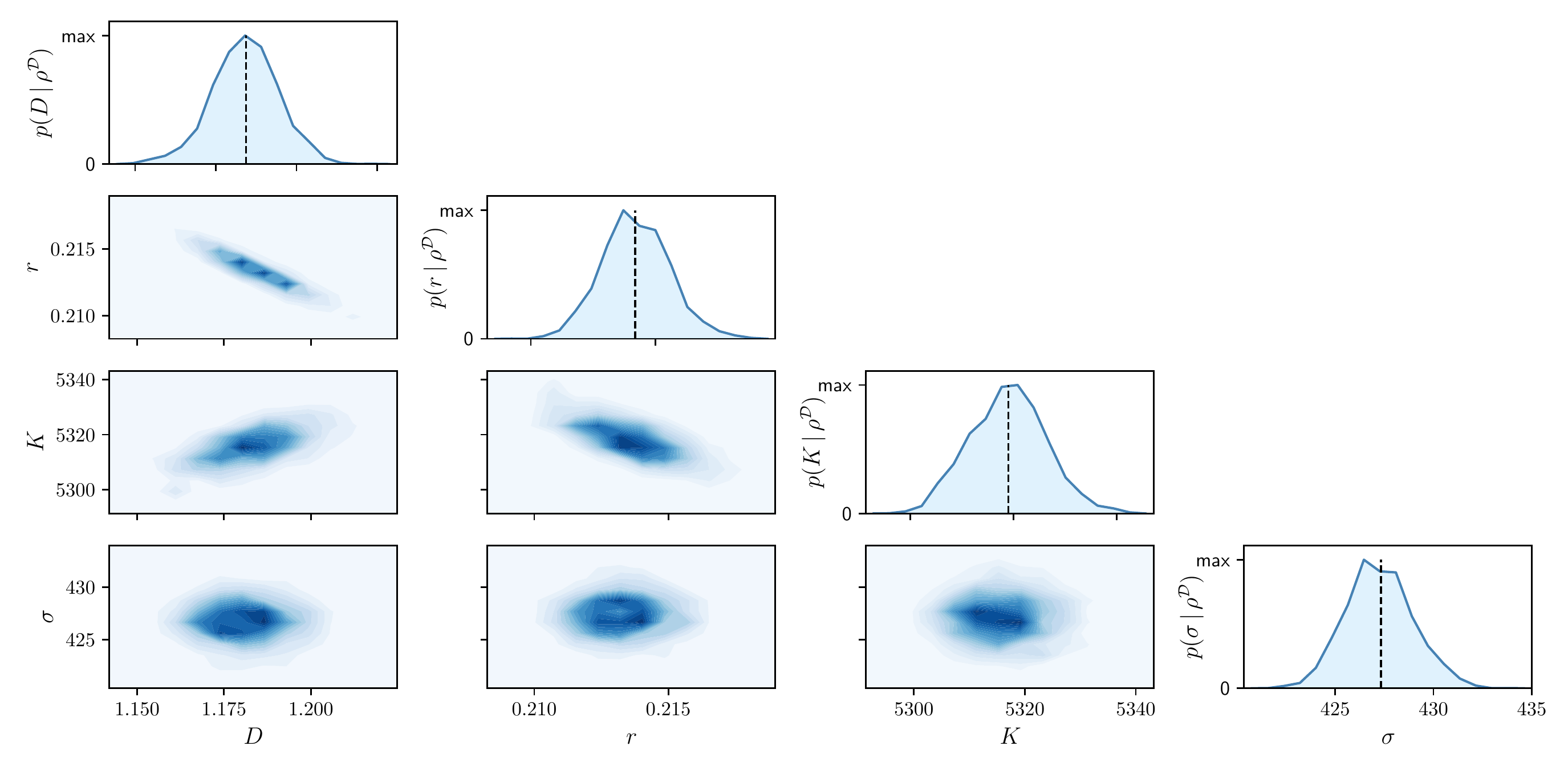}
        \caption{Results of the MCMC algorithm for the Porous-Fisher model given by Eq.~\eqref{eq:nonlinear_diffusion_1species}. The diagonal plots represent the univariate marginal posterior distributions for each parameter. Below the diagonal we show the bivariate densities for every combination of parameters. Univariate posterior modes correspond to $(D,r,K,\sigma) = (1.18\pm0.01$ $\mu$m$^2$/h, $0.214\pm0.001$ h$^{-1}$, $5319\pm7$ cells/mm$^2$, $427\pm2$ cells/mm$^2$),  where the errors are given by one standard deviation, calculated from the posterior distributions. Black dashed lines indicate the maximum likelihood estimates for each parameter.}
    \label{fig:posteriors_nonlinear}
\end{figure}

In order to infer the posterior distribution, we use a Metropolis-Hastings MCMC (Markov chain Monte Carlo) sampler with adaptive proposal covariance, which is also implemented in pyPESTO \cite{pypesto}. The Metropolis-Hastings MCMC algorithm  is a simple and popular choice for exploring the parameter space \cite{Hines2014DeterminationOP,simpson2020practical}, in which a Markov Chain starts at position $\theta$, and accepts a potential move to $\theta^*$ with probability $q = \min\{1,P(\theta\,| \,\rho^\mathcal{D})/P(\theta^*\,| \,\rho^\mathcal{D})\}$. In this way, the Markov chain tends to move towards high values of the posterior distribution, while still allowing for transitions to regions of lower probability in order to move away from local maxima. In this context, poor identifiability of the parameters can be detected by Markov chains that fail to converge towards a unimodal peaked posterior distribution.

We run the MCMC algorithm starting from three different initial guesses of $\theta$ for both models. In all cases, the Markov chains converge rapidly to narrow and well-defined stationary distributions --- see Figures S1 and S2 for plots of the chains and the univariate Gelman-Rubin convergence diagnostics. In particular, our typical Markov chain iterations are of length $12000$. Taking the last $5000$ iterations of the three chains in each model we obtain the posterior distributions $P(\theta\,| \,\rho^\mathcal{D})$. In Figures \ref{fig:posteriors_linear} (Fisher-KPP) and \ref{fig:posteriors_nonlinear} (Porous-Fisher) we show a plot matrix representation of the univariate and bivariate marginal distributions, with unimodal and approximately symmetric univariate densities. We also observe an excellent agreement between the marginal univariate modes and the maximum likelihood estimates found in the last section. Note that for the two models, different combinations of the parameters $D,r,K$ can result in the same invasion front speed, which explains the observed correlation between these parameters in the bivariate densities in Figures \ref{fig:posteriors_linear} and \ref{fig:posteriors_nonlinear}.  However, we observe that there is only one set of parameters maximising the likelihood, and that these parameters can be confidently identified given the small variance of the posterior distribution.

All identified parameters lie within the biologically feasible bounds. In the linear diffusion case (Fisher-KPP), the univariate modes are given by $(D,r,K,\sigma) = (1073$ $\mu$m$^2$/h, $0.29$ h$^{-1}$, $5113$ cells/mm$^2$, 492 cells/mm$^2$). Using an average density of $\sim 3000$ cells/mm$^2$ the estimated proliferation rate is around $\sim 0.1$ h$^{-1}$, which yields an estimated division time  around 10 hours. This is consistent with the characteristic division time for MDCK cells of 16-18 hours given that this timescale can vary significantly with cell size \cite{streichan2014spatial}. The carrying capacity can also be related to the typical cell radius for MDCK cells. Although notable variability has been reported \cite{zehnder2015cell}, the MDCK cell radius $a$ is estimated to oscillate between $a\sim6$ $\mu$m and $a\sim 18$ $\mu$m \cite{GauquelinMDCK}. Assuming that maximum densities in the monolayer are associated with hexagonal close packing of cells, the maximum theoretical density is given by $K = 1/(2\sqrt{3}a^2)$ \cite{vitadelloHexagonalPacking}. With our estimated carrying capacity this yields an estimate of $a\sim 8$ $\mu$m, which again is consistent with previous measurements, at least for cells in the bulk of the tissue.

In the case of the Porous-Fisher model, we obtain the univariate modes $(D,r,K,\sigma) = (1.18$ $\mu$m$^2$/h, $0.21$ h$^{-1}$, $5319$ cells/mm$^2$, 427 cells/mm$^2$). Note that the proliferation related parameters $r,K$ are very similar to the ones we estimated for the Fisher-KPP model. In this case, we estimate a cell divison time around $\sim 11$ hours, and a typical cell radius of $a\sim 7$ $\mu$m, again within the known ranges. Note that for the Porous-Fisher model, the diffusion coefficient is density-dependent --- $D(\rho) = D\rho$. Using an average density of $\sim 3000$ cells/mm$^2$, we also estimate an \emph{average diffusion coefficient}  which is three times larger than in the linear case, but still of the same order. We also observe that the estimated noise related parameter $\sigma$ is smaller in the Porous-Fisher case.

In summary, both models present well-defined and narrow posterior distributions for all the model parameters, with the parameter estimates being consistent with previous experimental measurements. Thus, we have shown via a Bayesian approach that all the model parameters appear to be identifiable. A more sophisticated approach aiming to use all the available data --- instead of measurements every 10 hours --- could include for instance a mini-batch algorithm \cite{martina2022efficient}. However, taking a subset of the data highlights that the models are practically identifiable, suggesting such approaches are not necessary in this case.

\subsection{Almost identical predictions from different continuum models}

Next, we explore to what extent the two considered models are able to reproduce the observed data. To do so, we solve numerically Eqs.~\eqref{eq:linear_diffusion_1species} and~\eqref{eq:nonlinear_diffusion_1species} using the parameter values that we estimated in the previous section. In order to minimise the possible impact of the stencil removal on cell motility \cite{heinrich2020size}, we use as initial condition the experimental density profile at time $t = 6$ h. The resulting radial density profiles are shown in Figure \ref{fig:expansions}C-D --- see also Figure S3.

First, we observe that both models yield very similar predictions with minor differences that are only noticeable near the expansion front. This is basically due to the fact that the solution of the Porous-Fisher model \eqref{eq:nonlinear_diffusion_1species} presents a sharp front, in contrast with the exponential decay in space of the Fisher-KPP equation \eqref{eq:linear_diffusion_1species}. Note that the Fisher-KPP model fails to accurately capture the behaviour of cell densities near the monolayer boundary, but the Porous-Fisher model, which accounts for population pressure, gives a more accurate description.

Secondly, we see that both models capture qualitatively the dynamics and growth of the expansions, but fail to capture the non-monotonic behaviour of the radial density profile for intermediate timescales. The experiments of Heinrich et al. \cite{heinrich2020size} observed that this phenomenon is accentuated for smaller tissues. Moreover, for later times, the experiments report cell densities that are higher than the estimated carrying capacity, which might be due to the fact that the considered models neglect variability in cell area as cells progress through their cell cycle \cite{khain2021dynamics,matsiaka2018discrete}.

All in all, these results show that both models, after being suitably calibrated, can explain equally well the data. As we will see next, it is only under more complex experimental conditions, when one needs to account for a more detailed level of physical description, that we can distinguish between the models.

\section{Quantifying tissue-tissue collisions}

Having seen that the two proposed models are practically identifiable, we now analyze how much mechanistic insight we can gain from more complex experiments. We consider a second set of experiments also performed by Heinrich et al. \cite{heinrich2021self}, where tissues are not isolated as in the previous experiments, but are allowed to interact with other tissues. In particular, Heinrich et al. study the dynamics of multi-tissue collisions, varying the shape and the number of colliding tissues, and find very complex patterns resulting from basic cell-cell interactions and mechanical properties. One of the most interesting observed features is the formation of sharp boundaries at the collision location, avoiding thus mixing of cells from different tissues, which is also characteristic of models that account for population pressure \cite{BertschPassoMimura,carrillo2018splitting}. Next, we follow closely these experiments and attempt to use both the Fisher-KPP and the Porous-Fisher models to reproduce different types of collisions.

Although we will always work with homotypic tissues (i.e. of the same cell type), it is particularly useful to identify a system consisting of multiple homotypic tissues with a model that accounts for several interacting cell populations. In our case, the tissues are composed of the same cell populations initially seeded at distinct spatial locations. Note, however, that the models presented below can account also for heterotypic tissue experiments. We denote the different species or tissues by $\rho_i$ for $i =1,\ldots,n$ with $n$ being the total number of species. In the linear diffusion Fisher-KPP model we assume that each species follows random motion and hence the diffusive part in the PDE remains unaffected. Taking into account that proliferation is limited by the total population density, we may write for $n = 2$
\begin{equation}
    \begin{cases}
    \partial_t\rho_1 = D\Delta\rho_1+r\rho_1\left(1 - \dfrac{\rho_1+\rho_2}{K}\right),\\
      \partial_t\rho_2 = D\Delta\rho_2+r\rho_2\left(1 - \dfrac{\rho_1+\rho_2}{K}\right).
    \end{cases}\label{eq:2species_linear}
\end{equation}

For the nonlinear diffusion Porous-Fisher model, we can write the total population pressure as $P(\rho_1,\rho_2) = D\left(\rho_1 + \rho_2\right)$. With this, the two-species model becomes
\begin{equation}
    \begin{cases}
         \partial_t\rho_1 = D\,\nabla\cdot\left(\rho_1 \nabla\left(\rho_1 + \rho_2\right)\right)+r\rho_1\left(1 - \dfrac{\rho_1+\rho_2}{K}\right),\\
      \partial_t\rho_2 = D\,\nabla\cdot\left(\rho_2 \nabla \left(\rho_1 + \rho_2\right)\right)+r\rho_2\left(1 - \dfrac{\rho_1 + \rho_2}{K}\right).
    \end{cases}
    \label{eq:2species_nonlinear}
\end{equation}
Extensions of these models to an arbitrary number of species, $n>2$,  are straightforward. The existence theory for cross-diffusion systems of the type of \eqref{eq:2species_nonlinear} is studied in \cite{burger2022porous,di2018nonlinear}. Note also that as a result of the population pressure term, system \eqref{eq:2species_nonlinear} gives sharp boundaries separating both species for initially segregated data \cite{BertschPassoMimura,carrillo2018splitting} which, again, motivates its use to reproduce the experiments in \cite{heinrich2021self}.

\subsection{Reproducing experimental tissue collisions}

In the next sections, we explore numerically the two proposed models under different initial conditions. We start with a qualitative study of some of the experiments performed by Heinrich et al. \cite{heinrich2021self} and follow with a more quantitative analysis of collisions between rectangular tissues.

\subsubsection{Simple binary tissue-tissue collisions}
\begin{figure}[h!t]
    \centering
    \includegraphics[width = \textwidth]{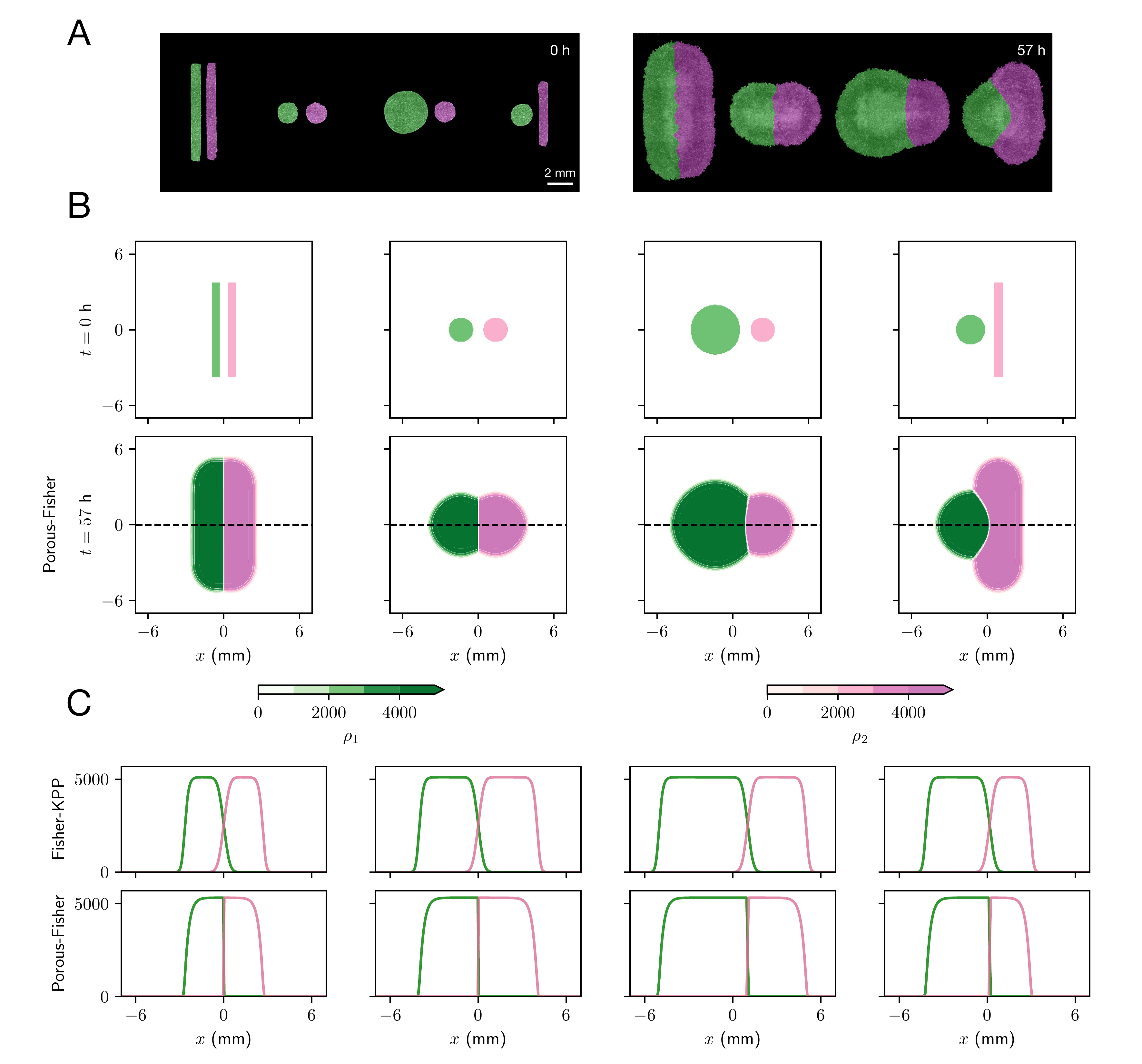}        \caption{Reproducing tissue-tissue collisions with different geometries --- animated movies available at \cite{figshare}. Accounting for population pressure correctly predicts the sharp boundaries observed in experiments. (A) Experimental results for initial conditions with different tissue geometries. Figures adapted from \cite{heinrich2021self} (\href{https://s100.copyright.com/AppDispatchServlet?title=Self-assembly\%20of\%20tessellated\%20tissue\%20sheets\%20by\%20expansion\%20and\%20collision&author=Matthew\%20A.\%20Heinrich\%20et\%20al&contentID=10.1038\%2Fs41467-022-31459-1&copyright=The\%20Author\%28s\%29&publication=2041-1723&publicationDate=2022-07-12&publisherName=SpringerNature&orderBeanReset=true&oa=CC\%20BY}{Creative Commons License}). Note that colours are only used to label each tissue and do not quantify cell densities. (B) Initial conditions and numerical simulations for the Porous-Fisher model (Eqs.~\eqref{eq:2species_nonlinear}) at $t = 57$ h. Colours in the numerical simulations indicate cell densities according to the shown colorbars. (C) Comparison of the Fisher-KPP model (Eq.~\eqref{eq:2species_linear}) and the Porous-Fisher model (Eq.~\eqref{eq:2species_nonlinear}). Solutions corresponding to the black dashed lines in (B). Parameter estimates given in the previous section: $(D,r,K) = (1073$ $\mu$m$^2$/h, $0.29$ h$^{-1}$, $5113$ cells/mm$^2)$ for the Fisher-KPP model and $(D,r,K) = (1.18$ $\mu$m$^2$/h, $0.21$ h$^{-1}$, $5319$ cells/mm)$^2$ for the Porous-Fisher model.}
    \label{fig:collisions}
\end{figure}

We first test the two proposed models in binary tissue-tissue collisions. In order to do so, here we choose different initial shapes for the two colliding tissues, namely we study circle-circle, rectangle-rectangle and circle-rectangle collisions. We also analyze the case of two colliding circles with different initial radii. See Figure \ref{fig:collisions}A for the experimental initial and final configurations. We emphasize that, in contrast with all the shown numerical simulations of our models, the colours in the experimental snapshots are only used to label each different tissue and do not quantify cell densities.

We numerically solve Eqs.~\eqref{eq:2species_nonlinear} for the four mentioned initial conditions and with the parameters that we estimated from the previous experiments \cite{heinrich2020size}. The numerical scheme is identical to the one-species case \cite{nonlocalNumericalSchemeRafaMarkus,nonlocalSchemceCarrilloYanghong}. As expected, in all four studied configurations the Porous-Fisher model shows sharp boundaries separating the two tissues after collision, and the observed patterns are nearly identical to the experimental final configurations after a simulation time equivalent to around 60 hours (Figure \ref{fig:collisions}B). Note that in contrast with the experimental snapshots, Figure \ref{fig:collisions}B shows quantitative cell densities.

\enlargethispage{.2cm}
When, instead of the Porous-Fisher model accounting for population pressure \eqref{eq:2species_nonlinear}, we use the Fisher-KPP model \eqref{eq:2species_linear}, we still observe patterns that resemble the experimental configurations. However, recall that in this case cells do not sense local pressure and are free to move in all directions, which results in a region where cells from both tissues can mix. Note that in this case, no sharp boundary between tissues is observed either --- Figure \ref{fig:collisions}C. Even though the Fisher-KPP model fails to reproduce density profiles near the collision boundary, it still can capture qualitatively the density profiles in the bulk of the tissue, where the population density gradient becomes more uniform. Hence, after suitable calibration, both the Fisher-KPP and the Porous-Fisher models show similar behaviour in this region far from the collision boundary and the propagating front.

Observe also that collisions shown in Figure \ref{fig:collisions}B that occur between tissues with the same shape (rectangle-rectangle and circle-circle collisions) were initialised with tissues of the same density. As reported experimentally in \cite{heinrich2021self}, these initial conditions result in the formation of a fixed sharp boundary that does not move in time. However, when collisions between tissues with different densities occur, then the denser tissue pushes the less dense tissue resulting in a boundary displacement which can be measured experimentally. For collisions between tissues with different shapes the collision boundary can also show a similar behaviour, as shown in Figure \ref{fig:collisions}. In the next sections we study this phenomenon quantitatively using the Porous-Fisher model \eqref{eq:2species_nonlinear}. Of course, given that linear diffusion fails to predict a sharp boundary between colliding tissues, this boundary displacement cannot be estimated from the Fisher-KPP model \eqref{eq:2species_linear}. Before moving to the study of collision boundary dynamics, we analyze a further set of more complex tissue collision experiments, which make evident the limitations of the simple Fisher-KPP model.

\subsubsection{Multi-tissue complex collisions}

In the previous sections we have showed that, after suitable calibration, both the Fisher-KPP and the Porous-Fisher models show similar behaviour in regions of tissue that are far from boundaries. However, under more complex experimental conditions where tissue boundary dynamics become important, the predictive power of the Fisher-KPP model becomes more limited.

\begin{figure}[t!]
    \centering
    \includegraphics[width = \textwidth]{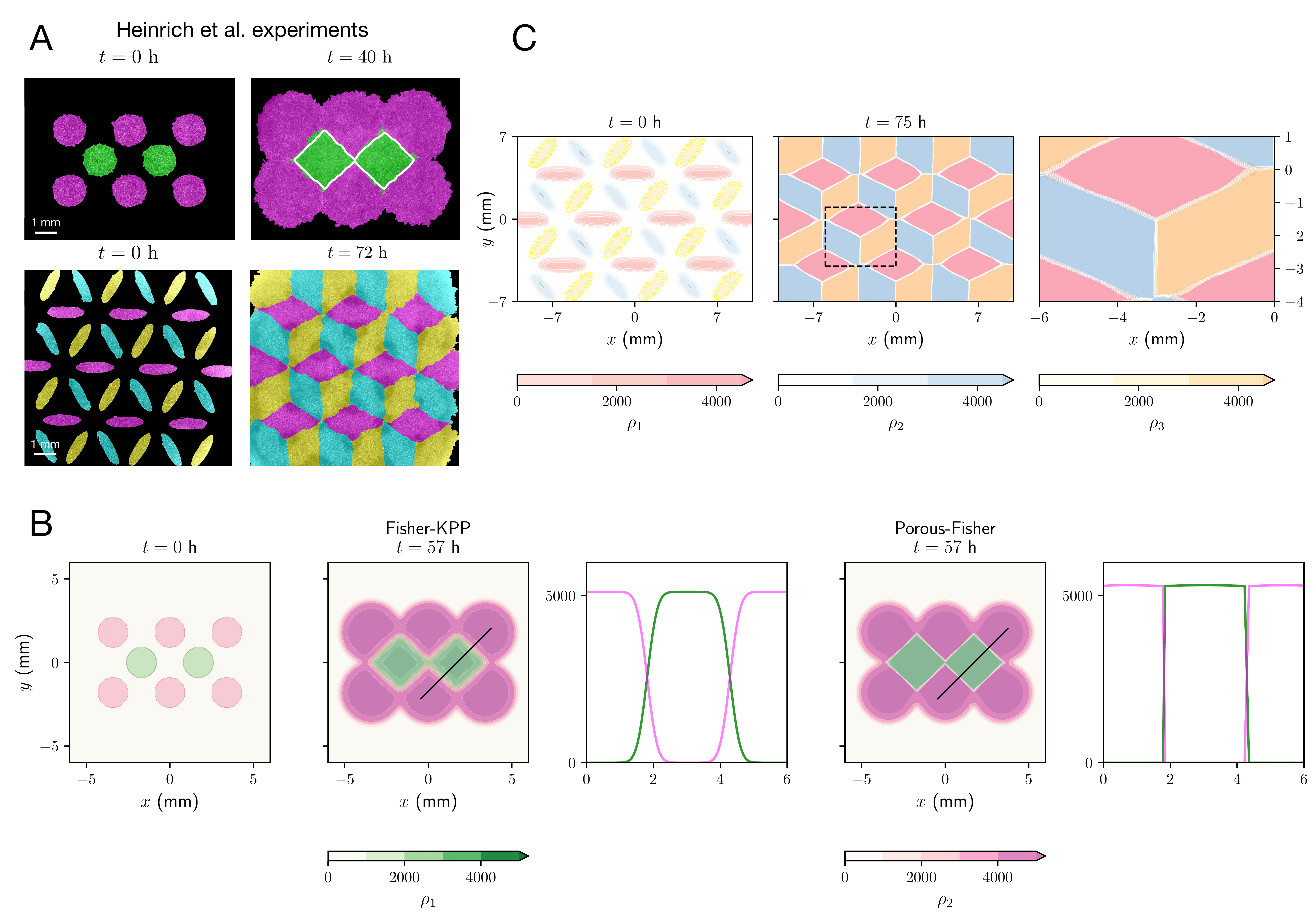}
    \caption{Reproducing complex tissue collisions observed in Heinrich et al. experiments --- animated movies available at \cite{figshare}. The Fisher-KPP model cannot reproduce complex multi-tissue collisions. (A) Experimental multi-tissue collisions, adapted from  \cite{heinrich2021self} (\href{https://s100.copyright.com/AppDispatchServlet?title=Self-assembly\%20of\%20tessellated\%20tissue\%20sheets\%20by\%20expansion\%20and\%20collision&author=Matthew\%20A.\%20Heinrich\%20et\%20al&contentID=10.1038\%2Fs41467-022-31459-1&copyright=The\%20Author\%28s\%29&publication=2041-1723&publicationDate=2022-07-12&publisherName=SpringerNature&orderBeanReset=true&oa=CC\%20BY}{Creative Commons License}) (B) Multi-tissue collision between eight homotypic circles for both the Fisher-KPP \eqref{eq:2species_linear} and the Porous-Fisher \eqref{eq:2species_nonlinear} models. Density profiles are taken along the black dashed lines. Note that numerical simulations use parameter estimates obtained from different experiments --- Figures \ref{fig:posteriors_linear} and \ref{fig:posteriors_nonlinear}. (C) Tri-tissue tesselation inspired by Escher's artwork and reproduced experimentally also by Heinrich et al \cite{heinrich2021self}. Here we show numerical simulations of the Porous-Fisher model. Rightmost panel zooms in the region indicated in the middle panel, and shows sharp boundaries. }
    \label{fig:collisions2}
\end{figure}

These differences between the Fisher-KPP \eqref{eq:2species_linear} and the Porous-Fisher model \eqref{eq:2species_nonlinear} become more evident when multiple tissues collide simultaneously. Here we focus on the experiments performed by Heinrich et al.  \cite{heinrich2021self} shown in Figure \ref{fig:collisions2}, where eight homotypic circular tissues are initially set apart on a hexagonal lattice. The initial configuration is also represented in Figure \ref{fig:collisions2} alongside the solutions predicted by the two proposed models after 57 hours. From these results, it becomes evident that the Fisher-KPP model is not suitable to describe complex interactions between tissues. In contrast, accounting for population pressure does yield the predicted behaviour with a final pattern nearly identical to that observed experimentally.

The Porous-Fisher model \eqref{eq:2species_nonlinear} can thus predict the behaviour observed in complex experimental settings with multiple tissues colliding. A numerical simulation of an extension of \eqref{eq:2species_nonlinear} to three species is depicted also in Figure \ref{fig:collisions2}. This last experiment mimics the self-assembly of a tri-tissue composite designed in \cite{heinrich2021self}.

\subsection{Quantifying collisions between rectangular tissues}

As mentioned earlier, collisions between two rectangular tissues result in the formation of a sharp boundary. Whenever the two rectangles are identical --- i.e. have the same shape and density --- the tissue boundary does not move and coincides with the centroid of the combined tissue. However,  Heinrich et al. observed that using larger or denser tissues results in a boundary displacement in the direction of the smaller or less dense tissue --- see Figure \ref{fig:collisions3}A for their experimental data. As shown in the Supplementary Information Section S2, the Porous-Fisher model also predicts this boundary displacement when there is a width/density mismatch between the initial tissues.

Here, we focus on the Porous-Fisher model, and explore to what extent it can reproduce the observed experimental data. In order to perform a quantitative comparison of model and experiments, we calibrate again Eqs.~\eqref{eq:2species_nonlinear} by using the data corresponding to a collision between identical rectangular tissues (control case in Figure \ref{fig:collisions3}A). After carrying out parameter estimation, we explore how the model performs in collisions of rectangular tissues with relative mismatches in either the width or number of cells (density and width mismatch in Figure \ref{fig:collisions3}A). For simplicity, and after having determined that our model is practically identifiable, we estimate the parameters using a maximum likelihood approach, as explained in previous sections, by comparing experimental and simulated cell densities. The initial densities are taken from experimental data, which in the control case (rectangles with equal density and equal width) are identical to those in Figure \ref{fig:collisions}A.

For this set of experiments, the maximum likelihood estimate yields $(D,r,K) = (3.26$ $\mu$m$^2$/h, $0.11$ h$^{-1}$, $4077$ cells/mm$^2$), which gives an approximate cell radius of $\sim 8$ $\mu$m. Observe that the diffusion parameter $D$ and the proliferation rate $r$ show notable differences with respect to the previous set of experiments. In particular these parameters suggest faster migration and slower proliferation, while the front speed remains more or less constant with respect to the case of a single tissue expansion. Note however, that as we proved in the previous sections, the Porous-Fisher model is practically identifiable and hence, although different parameter combinations result in the same invasion speed, we can confidently identify a set of parameters which maximises the likelihood of observing our data.

In fact, the differences in the parameters estimated from the two experiments \cite{heinrich2020size,heinrich2021self} could be explained by accounting for the transient regime that occurs immediately after the stencil removal. This short timescale is estimated to last around 6-8 hours, which we remove in order to calibrate the model. However, if we only take into account the first 20 hours of the experiment, the maximum likelihood procedure yields very different estimates for the model parameters, which suggests that the experimental collision time could be smaller than this transient timescale.

After the model is calibrated using the control case data, we can simulate Eqs.~\eqref{eq:2species_nonlinear} under different settings by varying the initial conditions. We study collisions between two rectangular tissues with an initial density (2600 vs 1800 cells/mm$^2$) or width (1000 vs 500 $\mu$m) mismatch. In Figure \ref{fig:collisions3}B we plot the density profiles obtained from the numerical simulations, which show an excellent agreement with the experimental data once both tissues have collided. At early times however, and in line with our discussion above, the model cannot reproduce the observed experimental dynamics. In particular tissue-tissue collisions occur around eight hours before they are observed in the experiments. The agreement between model and data becomes more evident upon visualising individual snapshots from these density profiles (see Figure \ref{fig:collisions3}C). Note that here, in the numerical simulation of both the density and width mismatch cases, we use the parameters estimated from a collision between identical rectangles.

\subsection{Population pressure gradients drive boundary displacement}

As discussed earlier, the Porous-Fisher model produces a sharp boundary separating the two colliding tissues. When the two tissues are not identical, there is a population pressure gradient at this boundary, which yields a net displacement with velocity $\mathbf{v} = - \nabla P(\rho)$. The nonlinear diffusion model assumes $P(\rho)\sim\rho$ and thus the boundary will move in the direction of the less crowded tissue. This translates, of course, into a wider tissue pushing a more narrow one, or a denser tissue pushing a less dense one.

\begin{figure}[t!]
    \centering
    \includegraphics[width = \textwidth]{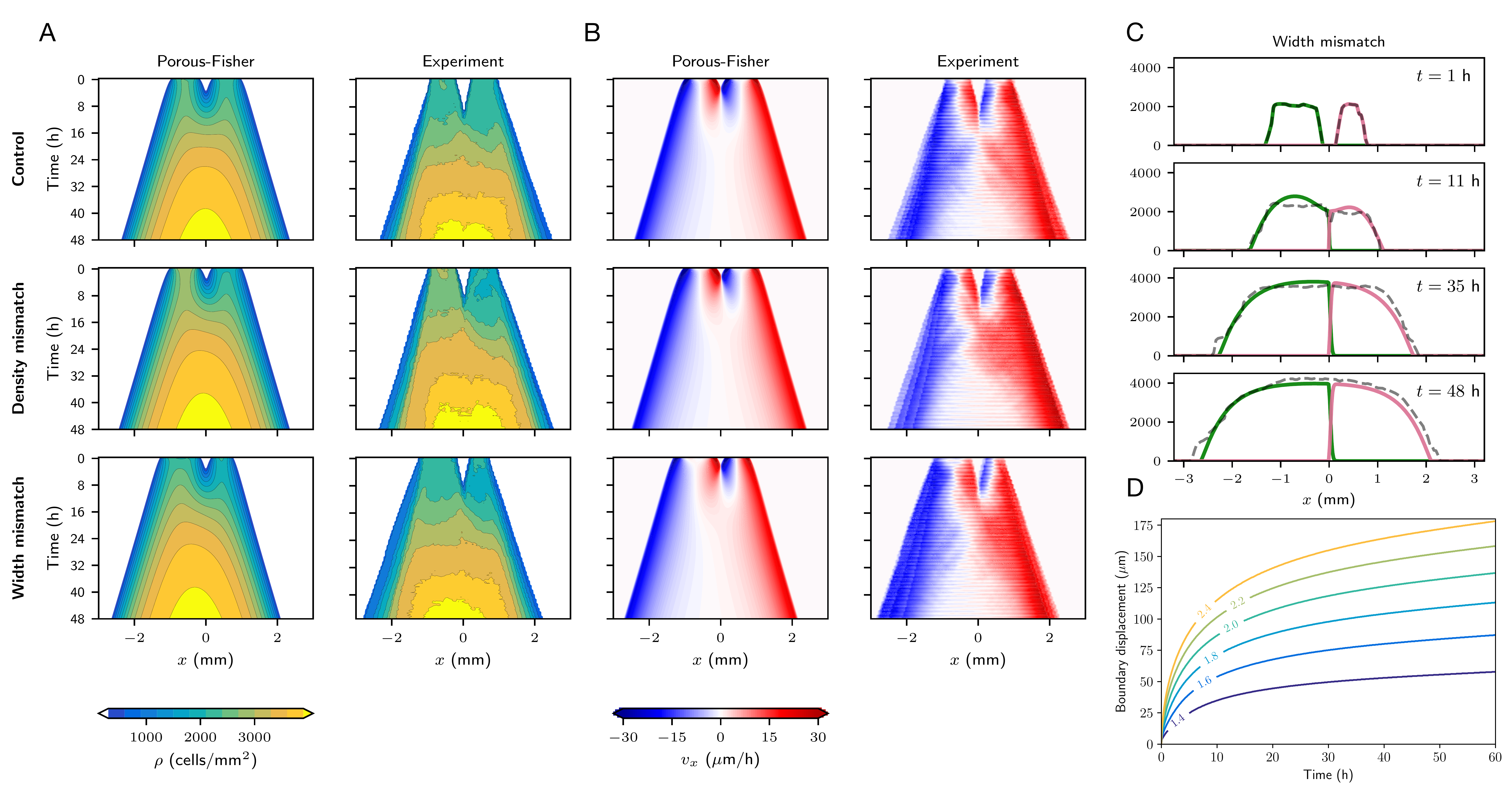}
    \caption{Quantifying rectangle-rectangle collisions. (A) Experimental density and velocity profiles resulting from rectangle-rectangle collisions --- data provided by Heinrich et al. \cite{heinrich2021self}. (B) Numerical simulations of the Porous-Fisher model \eqref{eq:2species_nonlinear} with parameters $(D,r,K) = (3.26$ $\mu$m$^2$/h, $0.11$ h$^{-1}$, $4077$ cells/mm$^2$). Initial conditions taken from the experimental data. (C) Comparing experimental density (dashed) with model prediction (solid) for the width mismatch case. Note that although parameters are estimated from the control case, there is an excellent agreement between model and data when the initial conditions are modified. (D) Boundary displacement predicted by the model, for different density mismatches, labelled in different colours.}
    \label{fig:collisions3}
\end{figure}

Our numerical simulations also reveal this behaviour (Figure \ref{fig:collisions3}D), giving a larger boundary displacement the larger the width/density mismatch. When using parameters inferred from the control case, the total boundary displacement that the model predicts falls short with respect to the experimental measures of Heinrich et al. by around 60-100 $\mu$m \cite{heinrich2021self}, which accounts for  less than $5\%$ of the final tissue width after collision. We believe that uncertainty associated with the experimental measures might have a minor impact on these results, as the boundary location can be determined experimentally up to subcellular accuracy and is then averaged over the collision axis --- given that different parts of the tissue might not collide at the same time. However,  the transient behaviour that cells exhibit after stencil removal can have a more significant effect on the later dynamics \cite{JIN2016136,jin2017logistic}, especially if this timescale is of the order of the collision time.

Another aspect which could have a more important influence on tissue boundary dynamics from the modelling perspective is the choice of the pressure function $P(\rho)$. In the Porous-Fisher model, cells move following population pressure gradients, moving away from crowded regions with a pressure function that is assumed to depend linearly on the density. However, using a more general pressure function would also give similar qualitative results but with possibly different dynamics. Note that a logarithmic dependence $P(\rho)\sim\log\rho$ \cite{heinrich2021self} is not suitable for this problem as it corresponds to the case of random cell movement in which there is no sharp boundary separating the tissues.

More generally, one could consider pressure functions that grow as a power-law function of the density, $P(\rho)\sim \rho^{m-1}$ for $m>1$. For large values of the exponent $m$, cells only move when the density gradient is large, while in the limit $m\rightarrow 1$ we recover the linear diffusion case. Considering this pressure-density relationship yields a porous-medium equation with proliferation for the evolution of the density, which also produces sharp boundaries between colliding tissues for $m>1$. Hence one could ask how does boundary displacement depend on the relationship between pressure and density --- i.e. on the exponent $m$. For small or no proliferation, this dependence can be analytically examined in the long-time regime. For instance, for one-dimensional tissues with an initial mass mismatch, a power-law pressure function yields a boundary displacement that grows in time as $\sim t^{1/(m-1)}$ thus giving a faster boundary motion for $m<1$ --- see Supplementary Information Section S2. We hence believe that it would be interesting to explore the practical identifiability of the exponent $m$, and whether considering a more general pressure-density relationship could give more accurate tissue boundary dynamics.

\enlargethispage{.45cm}

\section{Conclusion and outlook}

In this work, we have focused on two main aspects of tissue formation modelling: the practical identifiability of the Fisher-KPP and Porous-Fisher models using a Bayesian approach, and the applicability of the two models to describe tissue collisions experiments. Using data from recent experiments studying the growth and expansion of single epithelial sheets \cite{heinrich2020size}, we were able to obtain well-defined posterior distributions for each of the model parameters with relatively narrow confidence intervals. Our work thus adds to a growing literature assessing the practical identifiability of similar models under a variety of different experimental conditions \cite{browning2021model,simpson2020practical}.

In contrast with previous studies, and for the sake of conciseness, here we opted for using only a Bayesian MCMC approach. Another commonly used option is the profile likelihood method \cite{simpson2020practical,SIMPSON2022110998}, which  requires the solution of an optimization problem. This method however, can yield similar results to the MCMC algorithm and significantly reduce computational time. Although the Bayesian method can be very helpful in performing uncertainty quantification, we believe that studies comparing a larger number of models may benefit from a likelihood-based approach.

From a modelling perspective, we have proposed a systematic way to quantify cell densities and boundary locations in tissue collision experiments. This extends the model by Heinrich et al. \cite{heinrich2021self}, which was able to predict the boundary location for simple geometries and for tissues of the same initial density, but not to quantify tissue density. In contrast, our approach allows for more predictive power under a huge range of different experimental conditions. As discussed, being able to quantify and reproduce these tissue collision experiments is a first step towards the design and assemble of tissue composites.

This work could be extended by including other biological mechanisms in the models, such as more general pressure functions, cell-cell adhesion \cite{CarrilloMurakawaCellAdhesion,falco2022local}, cell-cycle dynamics \cite{heinrich2020size} or heterogeneity in cell size \cite{khain2021dynamics}, all of which  could improve our understanding of how cell interactions impact tissue collision dynamics. Although accounting for these different effects should be straightforward, whether the different model extensions are structural or even practically identifiable is not evident. Even simple models, very similar to the ones we used here, can lead to non-identifiability issues \cite{simpson2020practical}. Combining more detailed models with appropriate model selection and identifiability analysis thus seems challenging but also necessary in order to obtain better insights from the experimental work.

\vspace{1cm}

\textbf{Data access:} Experimental data used to calibrate our models (Figure \ref{fig:expansions}) is available on \cite{matthew_a_heinrich_2020_3858845}. Experimental data corresponding to tissue-tissue collisions (Figure \ref{fig:collisions3}) was provided by Heinrich et al. \cite{heinrich2021self}. Code used to perform the parameter estimation and to solve numerically the models is available on Github at: \url{https://github.com/carlesfalco/BInference-TissueCollisions}. Animated movies corresponding to the numerical simulations in the manuscript can be found on Figshare \cite{figshare} at: \url{https://figshare.com/projects/Quantifying_tissue_shape_growth_and_collision/157068}. Code used to create the animations is also available on Github.

\textbf{Acknowledgments:}
The authors thank M. Schmidtchen, M. A. Heinrich, A. E. Wolf for helpful discussions, and the members of the Hasenauer lab for providing assistance with the parameter estimation toolbox pyPESTO.

\textbf{Funding:}
JAC was supported by the Advanced Grant Nonlocal-CPD (Nonlocal PDEs for Complex Particle Dynamics: Phase Transitions, Patterns and Synchronization) of the European Research Council Executive Agency (ERC) under the European Union’s Horizon 2020 research and innovation programme (grant agreement No. 883363).
JAC was also partially supported by EPSRC grants EP/T022132/1 and EP/V051121/1. CF acknowledges support of a fellowship from "la Caixa" Foundation (ID 100010434) with code LCF/BQ/EU21/11890128.

\bibliographystyle{abbrv}
\bibliography{refs}

\end{document}


\section*{Supplementary Information}

\setcounter{figure}{0}
\renewcommand{\thefigure}{S\arabic{figure}}

\setcounter{section}{0}
\sectionfont{\fontsize{14}{15}\selectfont}
\renewcommand{\thesection}{S\arabic{section}}

\section{Supplementary Figures}

\begin{figure}[ht]
    \centering
    \includegraphics[width = \textwidth]{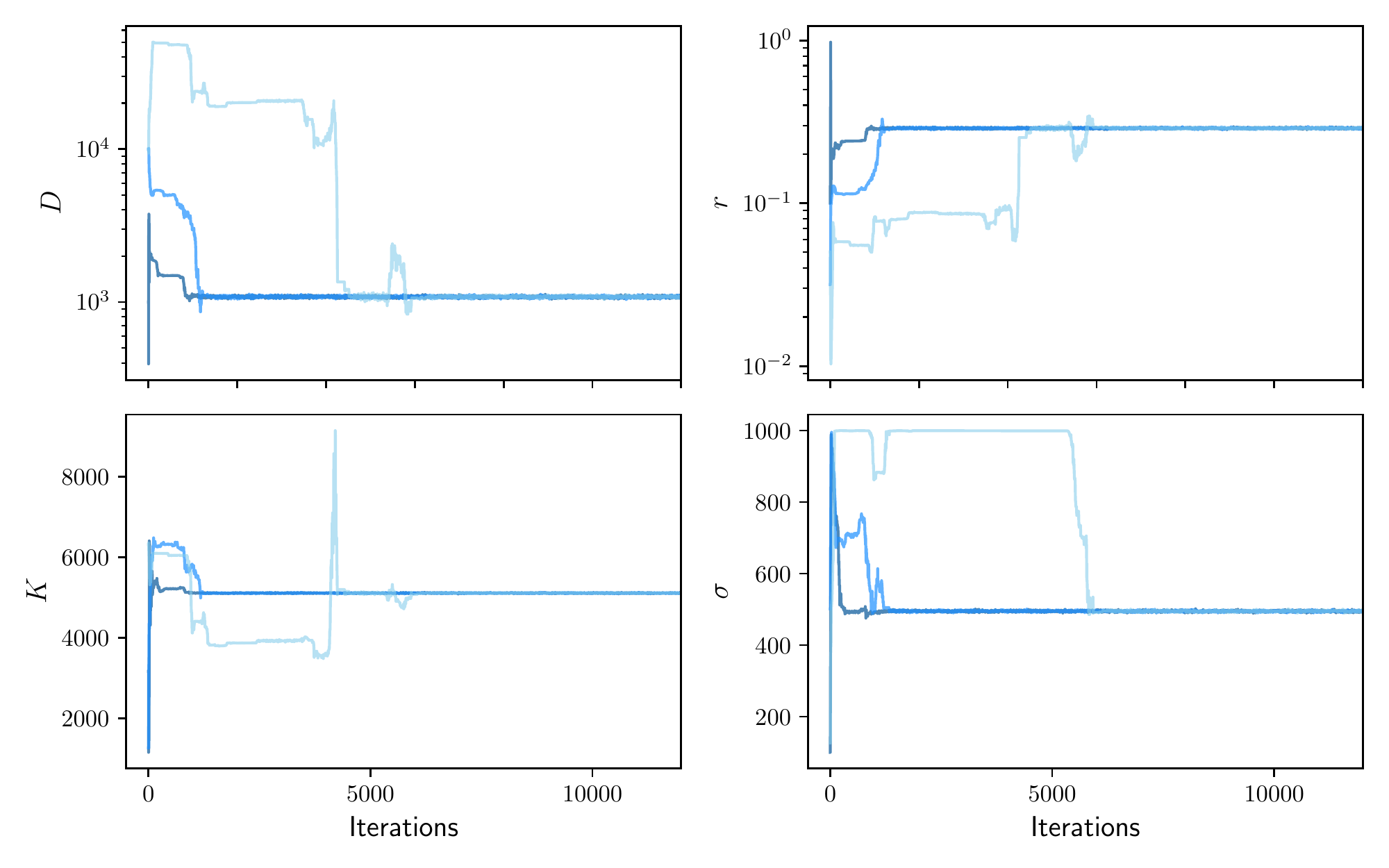}
    \caption{Typical Markov chain iterations of length $12000$ for the linear diffusion Fisher-KPP model and the parameters $D,r,K,\sigma$. The three shown chains are initiated with $(D,r,K,\sigma) = (10^3,0.1,3200,100), (10^4,0.03,1300,500), (10^4,0.05,6300,130)$ respectively. The maximum univariate Gelman-Rubin diagnostic among the four parameters satisfies $\hat{R}<1.01$ --- using the last 5000 chain iterations.}
    \label{fig:MCMC_chains_linear}
\end{figure}
\begin{figure}[ht]
    \centering
    \includegraphics[width = \textwidth]{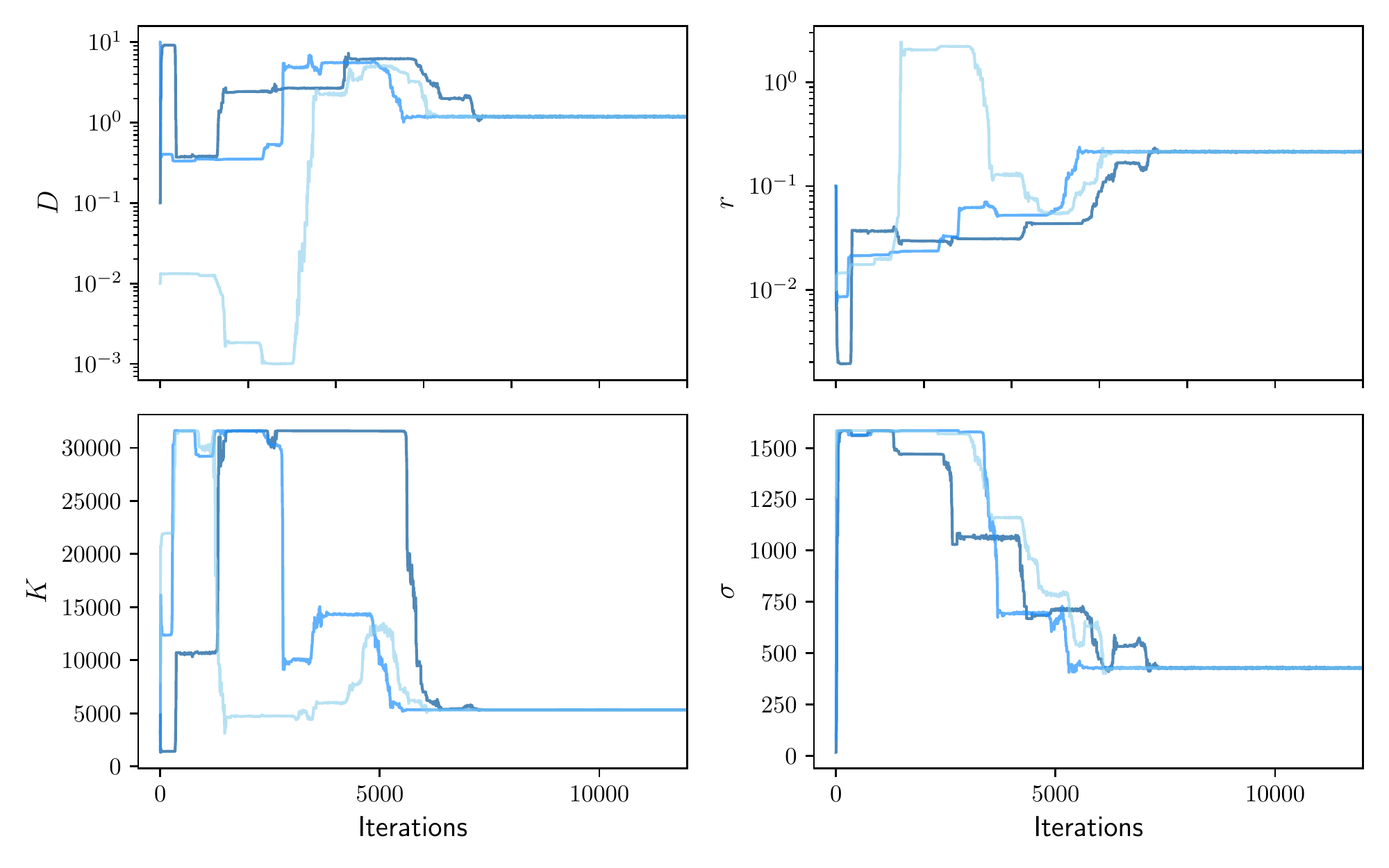}
   \caption{Typical Markov chain iterations of length $12000$ for the nonlinear diffusion Porous-Fisher model and the parameters $D,r,K,\sigma$. The three shown chains are initiated with $(D,r,K,\sigma) = (10^{-1},0.1,3200,160), (10,0.1,1600,100), (10^{-2},0.02,8000,120)$ respectively. The maximum univariate Gelman-Rubin diagnostic among the four parameters satisfies $\hat{R}<1.03$ --- using the last 5000 chain iterations.}
    \label{fig:MCMC_chains_nonlinear}
\end{figure}
\begin{figure}[ht]
    \centering
    \includegraphics[width = \textwidth]{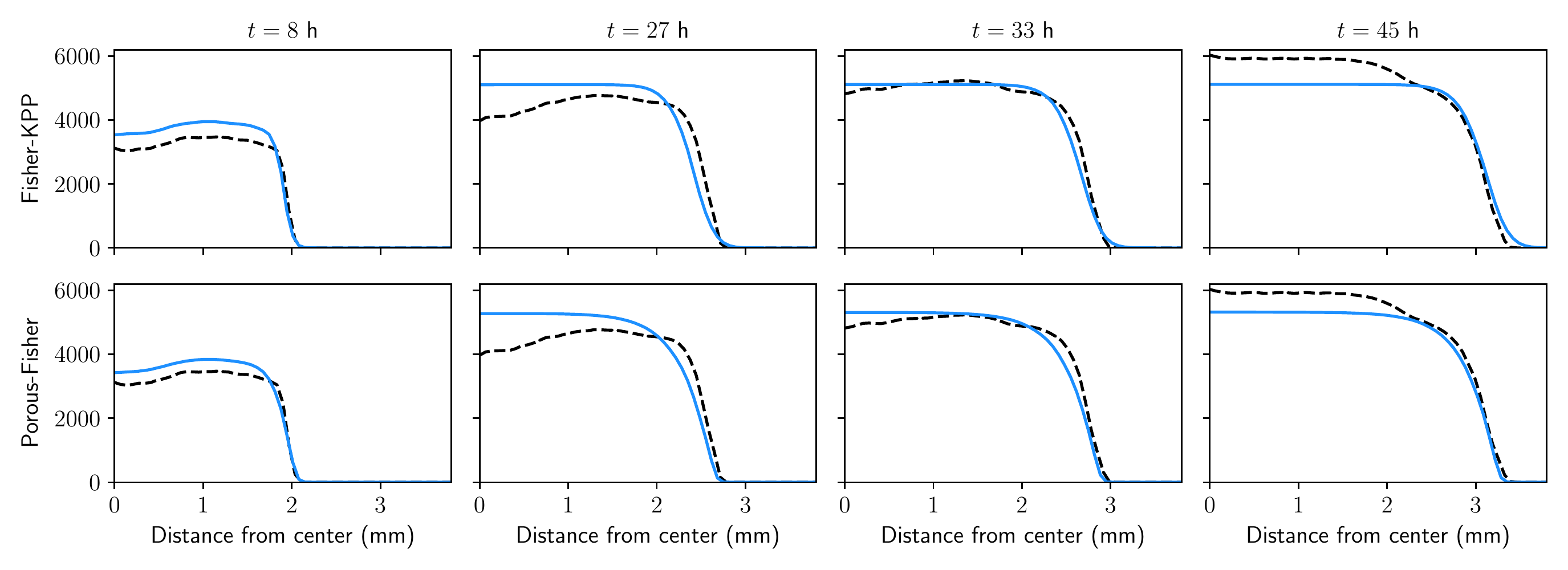}
   \caption{Comparing data and model prediction for tissue growth experiments \cite{heinrich2020size} --- individual snapshots corresponding to Figure 1 in the main text. Top row shows prediction from the Fisher-KPP (linear diffusion) model; bottom row shows results from the Porous-Fisher (nonlinear diffusion) model. Blue lines represent numerical simulations using the maximum likelihood estimate for the model parameters.}
    \label{fig:DataModelExpansions_SI}
\end{figure}

\clearpage
\section{Boundary displacement for cross-diffusion systems}
\label{sec: boundary displacement}

Here, we extend the Porous-Fisher model for two species to study analytically tisse boundary dynamics after a collision.

\subsection{Total population density and the porous medium equation}

We start by assuming that there is no proliferation, and hence the equations governing the dynamics of the two tissues read
\begin{equation*}
    \begin{cases}
        \partial_t \rho_1= D\,\nabla\cdot\left(\rho_1\nabla\left(\rho_1+\rho_2\right)\right),\\
       \partial_t \rho_2= D\,\nabla\cdot\left(\rho_2\nabla\left(\rho_1+\rho_2\right)\right).
    \end{cases}
\end{equation*}
The total population density, $\rho = \rho_1 + \rho_2$, satisfies then a porous medium equation with exponent two
\begin{equation*}
    \partial_t\rho = D\,\nabla\cdot\left(\rho\nabla\rho\right) = \frac{D}{2}\,\Delta\rho^2.
\end{equation*}
Note that this argument also works for more general versions of the population pressure, which in this case was assumed to grow linearly with the total population density: $P(\rho) \sim \rho$. More generally, if we assume that the pressure inside each tissue increases as a power-law function of the total population density, $P(\rho)\sim \rho^{m-1}$, we obtain the cross-diffusion system
\begin{equation*}
    \begin{cases}
        \partial_t \rho_1= D\,\nabla\cdot\left(\rho_1\left(\rho_1 + \rho_2\right)^{m-2}\nabla\left(\rho_1+\rho_2\right)\right),\\
      \partial_t \rho_2= D\,\nabla\cdot\left(\rho_2\left(\rho_1 + \rho_2\right)^{m-2}\nabla\left(\rho_1+\rho_2\right)\right).
    \end{cases}
\end{equation*}
The resulting equation for the total population density $\rho = \rho_1 + \rho_2$ is now a porous-medium equation with exponent $m$
\begin{equation*}
    \partial_t\rho =D \,\nabla\cdot\left(\rho^{m-1}\nabla\rho\right) =\frac{D}{m}\,\Delta\rho^m\,.
\end{equation*}

Solutions to the porous-medium equation in the whole space tend to the self-similar solution given by the Barenblatt profile \cite{carrillo2000asymptoticSI}, with known explicit expressions \cite{PMESI}.
For simplicity we deal with the one-dimensional case, although the argument works similarly in higher spatial dimensions. In that case, the solution\footnote{Assuming a proper rescaling of time and space to fix the constant in the porous medium equation} $\rho(x,t)$ tends as $t\rightarrow\infty$ to
\begin{equation*}
    \rho(x,t) = t^{-\alpha}\psi\left(|x|\, t^{-\alpha}\right),
\end{equation*}
where $\alpha = 1/(m+1)$ and
\begin{equation*}
    \psi(y) = \left(C-\kappa z^2\right)^{1/(m-1)},\quad |y| < \sqrt{\frac{C}{\kappa}},
\end{equation*}
with $\kappa = \alpha(m-1)/2D$ and the constant $C$ determined by conservation of mass. Note that $\rho$ propagates with a free boundary whose radius is given by $r(t) = \sqrt{C/\kappa}\, t^\alpha$. Imposing conservation of mass $M = \int_{-r(t)}^{r(t)}\rho(x,t)\, \mathrm{d}x$ yields an expression for the constant $C$
\begin{equation*}
    C^{1/({m-1})}\sqrt{\frac{C}{\kappa}}=\frac{\Gamma\left(\frac{1}{m-1}+\frac{3}{2}\right)}{\Gamma\left(\frac{m}{m-1}\right)}{\frac{M}{\sqrt{\pi}}}.
\end{equation*}

\subsection{Connecting pressure and boundary displacement}
We have seen that when the pressure is given by a power-law function of the density --- $P(\rho)\sim\rho^{m-1}$ --- the total population density $\rho$ is described by a porous-medium equation with exponent $m$, whose asymptotic solution is known. Further, we also know that due to the population pressure term, the two species do not mix but stay segregated \cite{BertschPassoMimuraSI,carrillo2018splittingSI}. This means that there is an interface separating them whose position is given by $b(t)$. In order to find the position of this interface we just need to impose mass conservation for one of the tissues: $M_2 = \int_{b(t)}^{r(t)}\sigma(x,t)\,\mathrm{d}x$ --- where the masses of the two colliding tissues are given by $M_1$ and $M_2$. By doing so, we obtain
\begin{align}
    M_2 = \int_{b(t)}^{r(t)}\sigma(x,t)\,\mathrm{d}x &= t^{-\alpha}C^{1/(m-1)}\int_{b(t)}^{r(t)}\left(1 - \frac{\kappa x^2}{C t^{2\alpha}}\right)^{1/(m-1)}\mathrm{d}x \nonumber\\&= \frac{\Gamma\left(\frac{1}{m-1}+\frac{3}{2}\right)}{\Gamma\left(\frac{m}{m-1}\right)}{\frac{M}{\sqrt{\pi}}}\int_{b(t)/r(t)}^1 \left(1 - y^2\right)^{1/(m-1)}\,\mathrm{d}y\,.\label{eq:SI_mass 2}
\end{align}
Eq. \eqref{eq:SI_mass 2} predicts the behaviour of $b(t)$. By noting that the integral cannot depend on time, we obtain that $b(t)$ is a fraction of the expansion radius, $b(t)/r(t) = \ell < 1$, and hence $b(t)$ grows at the same rate as $r(t)$ \begin{equation*}
    b(t)\sim t^\alpha,\quad \alpha = \frac{1}{m+1}\,.
\end{equation*}
Note that this equation relates the form of the equation of state --- $P(\rho)\sim \rho^{m-1}$ --- with the boundary displacement, which can be measured experimentally.

The ratio $\ell$ can be found by solving the relation
\begin{equation*}
    \frac{2M_2}{M} = \frac{\int_\ell^1\left(1 - y^2\right)^{1/(m-1)}\,\mathrm{d}y}{\int_0^1\left(1 - y^2\right)^{1/(m-1)}\,\mathrm{d}y}\,,
\end{equation*}
or in terms of the Beta function
\begin{equation*}
    \frac{M_1 - M_2}{M} = \frac{\mathrm{B}\left(\ell^2;1/2,\frac{m}{m-1}\right)}{\mathrm{B}\left(1/2,\frac{m}{m-1}\right)},
\end{equation*}
where $B(\ell^2\,;a,b)$ is the incomplete Beta function evaluated at $\ell^2$.
It is clear then that if $M_2 = M_1 = M/2$, then $\ell = 0$ and that $M_2 < M_1$ (density or width mismatch) allows for boundary displacement with $0 < \ell < 1$.
\subsection{Boundary displacement in the presence of proliferation}
In a setting where the cell divison time is smaller than the timescale of the experiment one can consider a system equivalent to the ones in the previous section but with a linear proliferation term
\begin{equation*}
\begin{cases}
    \partial_t \rho_1= D\,\nabla\cdot\left(\rho_1\left(\rho_1 + \rho_2\right)^{m-2}\nabla\left(\rho_1+\rho_2\right)\right) + r\rho_1\,,\\
    \partial_t \rho_2= D\,\nabla\cdot\left(\rho_2\left(\rho_1 + \rho_2\right)^{m-2}\nabla\left(\rho_1+\rho_2\right)\right)+r\rho_2.
\end{cases}
\end{equation*}
Now the equation for the total population density $\rho = \rho_1 + \rho_2$ reads
\begin{equation*}
    \partial_t\rho = \frac{D}{m}\,\Delta\rho^m + r\rho.
\end{equation*}
Under the change of variables  ${\rho}= \Tilde{\rho} e^{r t}$ and $\tau = ({e^{r(m-1)t}-1})/{r(m-1)}$ \cite{gurtin1977diffusionSI}, the equation for the total population density reduces to  \begin{equation*}
    \partial_\tau\Tilde{\rho} =\frac{D}{m}\,\Delta\Tilde{\rho}^m.
\end{equation*}
Now the boundary satisfies
\begin{equation*}
    b(t) \sim \tau^\alpha \sim \left(e^{r(m-1)t}-1\right)^{\alpha},
\end{equation*}
which shows the same power-law behaviour for $t\ll 1/r(m-1)$, but grows exponentially for $t>1/r(m-1)$.

Note however that the exponential growth approximation eventually becomes biologically unrealistic as at some point proliferation will become density-limited. At this point, the densities at both tissues are uniform, driving the boundary to slow down until it finally stops moving.

\bibliographystyle{abbrv}
\bibliography{refs2}